# All-magnetic control of skyrmions in nanowires by a spin wave

Xichao Zhang[1], Motohiko Ezawa[2*], Dun Xiao[3], G. P. Zhao[4, 5], Yaowen Liu[3] and Yan Zhou[1†]

[1] *Department of Physics, The University of Hong Kong, Hong Kong, China*

[2] *Department of Applied Physics, University of Tokyo, Hongo 7-3-1, 113-8656, Japan*

[3] *Shanghai Key Laboratory of Special Artificial Microstructure Materials and Technology, School of Physical Science and Engineering, Tongji University, Shanghai 200092, China*

[4] *College of Physics and Electronic Engineering, Sichuan Normal University, Chengdu 610068, China*

[5] *Key Laboratory of Magnetic Materials and Devices, Ningbo Institute of Material Technology & Engineering, Chinese Academy of Sciences, Ningbo 315201, China*

\* E-mail: ezawa@ap.t.u-tokyo.ac.jp

† E-mail: yanzhou@hku.hk

**Abstract**

Magnetic skyrmions are topologically protected nanoscale objects, which are promising building blocks for novel magnetic and spintronic devices. Here, we investigate the dynamics of a skyrmion driven by a spin wave in a magnetic nanowire. It is found that (i) the skyrmion is first accelerated and then decelerated exponentially; (ii) it can turn L-corners with both right and left turns; and (iii) it always turns left (right) when the skyrmion number is positive (negative) in the T- and Y-junctions. Our results will be the basis of skyrmionic devices driven by a spin wave.

Keywords: skyrmion, spin wave, Dzyaloshinskii-Moriya interaction, Thiele equation

**1. Introduction**
　　The concept of topology has sparked wide interest in recent years. A well-known example is the quantum spin Hall edge state of the topological insulator [1-3], which is protected by the time-reversal symmetry and therefore has attracted tremendous interests in condensed matter physics. A popular example in topology is that both a coffee mug and a donut can continuously transform into a torus, indicating the same nature of topology for the donut and coffee mug. On the contrary, a coffee mug (or donut) cannot morph into a sphere without introducing rupture, meaning they are of different topology. A significant amount of energy is required in order to transform a certain object into another with different topology, which could be described as "topological stability" or "topological protection". Topological considerations are of considerable use in describing and understanding the extraordinary stability of such system. The concept of topology is also of crucial importance in studying liquid crystals, vortex in superconductors, and superfluids *etc.* in condensed matter systems [4].

　　The concept of skyrmion was firstly proposed by Tony Skyrme to describe the interactions of pions in the context of nuclear physics [5-8]. Later it is generalized to various subjects in condensed matter physics including quantum Hall magnets, Bose-Einstein condensate *etc.* [9]. A skyrmion is a topological particle-like excitation in classical continuum field theory which is robust as long as the field is continuous and the edge effect is negligible. In magnetic materials, a wide range of magnetic configurations are being researched intensively in the form of domain walls, vortices, monopoles, and magnetic skyrmions in recent years due to the same topological concern [4]. In analogy to the well-known example of donut-balloon transformation, a magnetic skyrmion cannot be continuously transformed into other magnetic configurations such as ferromagnetic state, without surpassing the topological energy barrier. Therefore, magnetic skyrmion is topologically protected and relatively more stable than other types of magnetic configurations such as vortex and bubble, making it very promising for realistic applications in information processing and ultra-high density information storage [9, 10]. Recent experimental realizations of skyrmions in magnet have attracted great interest [9-21]. Most of the experimental observations of skyrmions are reported in non-centrosymmetric ferromagnets such as MnSi, FeGe and $Fe_{0.5}Co_{0.5}Si$ *etc.* [10-19, 21], *i.e.*, the B20-type materials. More recently, isolated skyrmion was also successfully realized in thin films of similar materials lacking inverse symmetry or in proximity of heavy metal substrate inducing sizable DMI [22-24]. A skyrmion can be created by circulating current [25], from notch [26], from photo-irradiation [27, 28], or from a domain-wall pair [29]. Skyrmion is expected to be a key player of the next-generation electronics – skyrmionics [9, 10]. Skyrmion can be driven by spin-polarized current [30-34]. However, to move the skyrmion along the central line of a nanotrack by in-plane spin-polarized current requires severe matching between the damping coefficient and the non-adiabatic coefficient, *i.e.* $\alpha$ is close to $\beta$, limiting possible material systems for skyrmion applications [9, 26, 29, 32]. Another possibility of controlling a skyrmion is to use spin wave [34, 35]. Spin wave produces less heat than electric current, which therefore is promising for practical applications [36-39]. We investigate the conditions permitting one to use spin waves, instead of electrical currents, to control skyrmions in nano-circuits. So far there is no report on the skyrmion dynamics driven by spin wave in constricted geometries such as nanotrack and junction. In a real skyrmionic device, skyrmions will travel in circuits consisting of narrow nanotracks. Thus the study of the skyrmion dynamics in such configurations is crucial for realization of skyrmionics.

　　In this paper, we investigate the skyrmion dynamics driven by spin wave in constricted geometries with the Dzyaloshinskii-Moriya interaction (DMI) such as nanotracks, L-corners, T- and Y-junctions, which are the basic ingredients of circuits based on skyrmions. Our major findings are as follows: 1) A skyrmion can travel quite a long distance without touching sample edges and without requiring fine-tuning of sample parameters. 2) We show that a skyrmion can turn a shaped corner without touching the sample edges even in the case of the L-corner. 3) A skyrmion always turns left (right) at the T- or Y-junctions when the topological number is positive (negative). 4) The turning direction of the skyrmion at the junction can be controlled by using multiple spin wave injection sources. 5) The skyrmion velocity can be well explained by a fitting function which embodies its initial acceleration and subsequent exponential decay.

**2. Methods**
2.1. Simulation details
The micromagnetic simulations are performed using the Object Oriented MicroMagnetic Framework (OOMMF) [40] including the DMI module [41-43]. The time-dependent magnetization dynamics is governed by the

Landau-Lifshitz-Gilbert (LLG) equation [44-47]

$$\frac{d\mathbf{M}}{dt} = -|\gamma|\mathbf{M} \times \mathbf{H}_{\text{eff}} + \frac{\alpha}{M_S}\left(\mathbf{M} \times \frac{d\mathbf{M}}{dt}\right), \quad (1)$$

where $\mathbf{M}$ is the magnetization, $\mathbf{H}_{\text{eff}}$ is the effective field, $\gamma$ is the Gilbert gyromagnetic ratio, and $\alpha$ is the damping coefficient. The effective field is defined as follows:

$$\mathbf{H}_{\text{eff}} = -\mu_0^{-1} \frac{\partial E}{\partial \mathbf{M}}. \quad (2)$$

The average energy density $E$ is a function of $\mathbf{M}$ specified by [32, 44, 48],

$$E = A\left[\nabla\left(\frac{\mathbf{M}}{M_S}\right)\right]^2 - K\frac{(\mathbf{n}\cdot\mathbf{M})^2}{M_S^2} - \mu_0 \mathbf{M}\cdot\mathbf{H} - \frac{\mu_0}{2}\mathbf{M}\cdot\mathbf{H}_d(\mathbf{M}) + \frac{D}{M_S^2}\left(M_z\frac{\partial M_x}{\partial x} - M_x\frac{\partial M_z}{\partial x} + M_z\frac{\partial M_y}{\partial y} - M_y\frac{\partial M_z}{\partial y}\right), \quad (3)$$

where $A$ and $K$ are the exchange and anisotropy energy constants, respectively. $\mathbf{H}$ and $\mathbf{H}_d(\mathbf{M})$ are the applied and magnetostatic self-interaction fields while $M_S = |\mathbf{M}(\mathbf{r})|$ is the spontaneous magnetization. $D$ is the DMI constant and $M_x$, $M_y$, $M_z$ are the components of the magnetization $\mathbf{M}$. The five terms at the right side of Eq. (3) correspond to the exchange energy, the anisotropy energy, the applied field (Zeeman) energy, the magnetostatic (demagnetization) energy and the DMI energy, respectively.

For micromagnetic simulations, we consider 1-nm-thick cobalt nanotracks on the substrate with low damping [33, 49, 50]. The intrinsic magnetic parameters are similar to Ref. [32]: Gilbert damping coefficient $\alpha = 0.01 \sim 0.05$ and the value for $\gamma$ is $-2.211\times10^5$ m A$^{-1}$ s$^{-1}$. Saturation magnetization $M_S = 580$ kA m$^{-1}$, exchange stiffness $A = 15$ pJ m$^{-1}$, DMI constant $D = 4$ mJ m$^{-2}$ and perpendicular magnetic anisotropy (PMA) $K = 0.8$ MJ m$^{-3}$ unless otherwise specified. Thus, the exchange length is $l_{\text{ex}} = \sqrt{\frac{A}{K}} = 4.3$ nm. The simulated models are discretized into $2 \times 2 \times 1$ nm$^3$ cells except the Y-junctions, which are discretized into $1 \times 1 \times 1$ nm$^3$ cells in order to maintain the numerical accuracy.

In the simulation of SW-driven skyrmion in the nanotrack, the width (along $y$) of the nanotrack is 40 nm and the length (along $x$) is 800 ~ 1500 nm. The width of the pulse element equals to the width of the nanotrack and the length is fixed to be 15 nm. In the simulation of SW-driven skyrmion in the T- and Y-junction, the width of the nanotrack is increased to 60 nm and $D$ is decreased to 3.5 mJ m$^{-2}$, which broadens the channel and reduces the size of the skyrmion, leading to a better effect of the skyrmion turning at the junction.

The virgin state of the magnetization of the nanotrack is relaxed along $+z$ direction, except for the tilted magnetization near the edges due to the DMI. At first, a skyrmion is created at designated spot (as shown in Fig. 1) by the vertical spin-polarized current injection [32] and relaxed to stable/metastable state within a short period of time. We also implement absorbing boundary conditions (ABCs) based on an exponential increase of the damping coefficient at the ends of the nanotrack to eliminate any abrupt changes in damping and effectively suppress any spurious spin wave reflections [51].

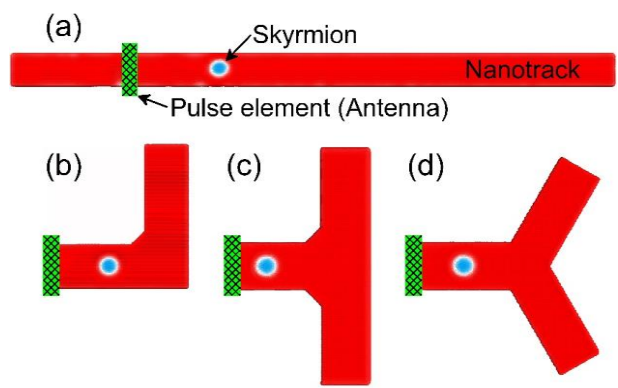

Figure 1. Schematics of the micromagnetically modeled system. (a) The magnetic nanotrack. (b) The L-corner. (c) The T-junction. (d) The Y-junction with angle of 120°. The patterned green boxes denote the pulse elements, *i.e.*, the microwave antenna placed upon the nanotrack, from where the spin wave is injected via the applied magnetic pulse applied along the lateral axis of the nanotrack.

## 3. Results and discussion
### 3.1. A skyrmion on nanotrack
We first demonstrate the spin wave driven motion of the skyrmion on an 800-nm-long and 40-nm-wide nanotrack with the spin wave (SW) injection.

For the SW injection as shown in Fig. 1(a), a skyrmion is located at $x = 200$ nm at $t = 0$ ns. A magnetic field pulse is applied by the pulse element on the left side of the track (135 nm < $x$ < 150 nm), which can be realized by employing a microwave antenna placed upon the nanotrack [31, 52, 53]. The profile of the square magnetic field pulse is shown in the inset of Fig. 3. The amplitude of the field is 600 mT and both the pulse width and spacing are 0.02 ns, i.e., the frequency is 25 GHz. The magnetic field pulse is applied perpendicularly to the track (see Supplementary Note 1 for the parallel case). The excited SW propagates toward the ends of the nanotrack and drives the skyrmion into motion at the same time. Figure 2(a) shows the propagation of the skyrmion driven by SW in the 40-nm-wide nanotrack with the damping coefficient of 0.01. At $t = 5$ ns, the skyrmion moves 266 nm along the nanotrack with an average speed of 53 m s$^{-1}$. At $t = 9$ ns, it moves 457 nm along the nanotrack with an average speed of 51 m s$^{-1}$ (see Supplementary Movie 1 and Supplementary Note 2). However, when the damping coefficient of the nanotrack increases to 0.02, the skyrmion moves 180 nm along the nanotrack with an average speed of 20 m s$^{-1}$ at $t = 9$ ns. When the damping coefficient further increases to 0.05, the skyrmion only moves 49 nm along the nanotrack with an average speed of 5 m s$^{-1}$ at $t = 9$ ns, as shown in Fig. 2(b) and 2(c). Hence, it can be seen that the mobility of the skyrmion on the nanotrack reduces significantly with increasing damping coefficient of the system under the same condition of SW injection, since the excited spin wave decays quickly as its amplitude decreases with the damping coefficient $\alpha$. For this reason, in order to show a better performance of SW-driven skyrmion, the damping coefficient in all simulations is fixed at 0.01 in the remaining of this paper. However, for large damping constant results, please refer to the Supplementary Information Note 3.

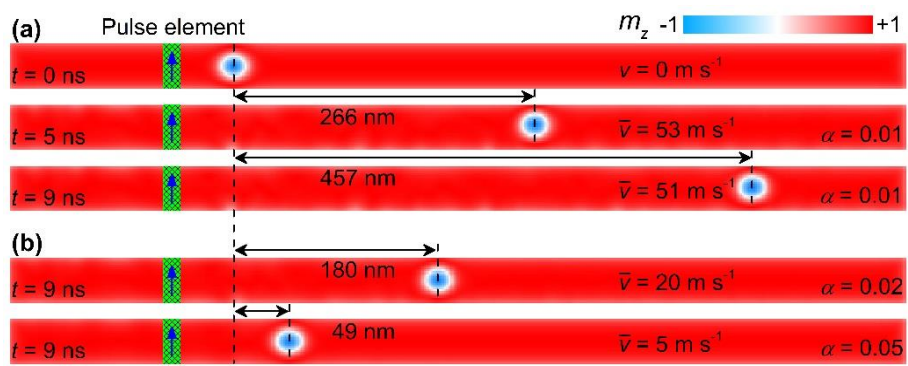

Figure 2. The propagation of a skyrmion driven by the spin wave in the 40-nm-wide nanotrack. The patterned green boxes on the track corresponds to the region of the spin wave injection (135 nm < $x$ < 150 nm). (a) Snapshots of the propagation of the skyrmion on the nanotrack with the damping coefficient of 0.01 (also see Supplementary Figure 3 for cross-sectional views). (b) Snapshots of the propagation of the skyrmion on the nanotracks with larger damping coefficients of 0.02 and 0.05. The color scale presents the out-of-plane component of the magnetization m$_z$, which has been used throughout this paper.

Figure 3 shows the skyrmion's velocity as a function of time in 800-nm-long and 1500-nm-long nanotracks. Obviously, on the 800-nm-long nanotrack, the skyrmion's velocity is not uniform and experiences acceleration and deceleration. The skyrmion reaches the end of the nanotrack at $t \sim 13.5$ ns. The skyrmion slows down due to the skyrmion-edge repulsion and finally it stops at a position balanced by the skyrmion-edge repulsive force and the SW driving force. During the first 9 ns, the maximal velocity is $\sim 67$ m s$^{-1}$, and the average velocity is $\sim 51$ m s$^{-1}$. In the 1500-nm-long nanotrack, the skyrmion are far away from both the end of the nanotrack and the source of the SW at $t = 20$ ns, resulting in the exponential decrease of its velocity. It should be mentioned that we also investigated the case where the magnetic field pulse is parallel to the track instead of perpendicular to it, which shows similar results (see Supplementary Note 1).

Figure 4 shows the skyrmion's velocity as a function of time by varying different parameters with the magnetic field pulse as shown in the inset of Fig. 3. A larger average and maximal speed can be achieved for the skyrmion in the nanotrack with larger DMI strength $D$, smaller perpendicular magnetic anisotropy $K$ and smaller exchange stiffness $A$. For the nanotrack with larger $D$ and smaller $K$, the equilibrium size of the skyrmion is larger, leading to a larger surface of the skyrmion to interact with the SW, which results in larger driving force. For larger $K$, the spins are harder to flip and the SW decays faster, resulting in a smaller velocity of the skyrmion. For smaller $A$, the spins around the skyrmion are easier to be reversed, leading to a larger velocity. As shown in Fig. 4(d), the average/maximal skyrmion velocity increases by decreasing $M_S$ from 620 kA m$^{-1}$ to 580 kA m$^{-1}$. However, when $M_S$ reduces below 560 kA m$^{-1}$, the average/maximal skyrmion velocity dramatically decreases. Form Fig. 4(e) and Fig. 4(f), it can be seen that the larger amplitude and/or higher frequency of the magnetic pulse lead to a larger average/maximal speed of the skyrmion.

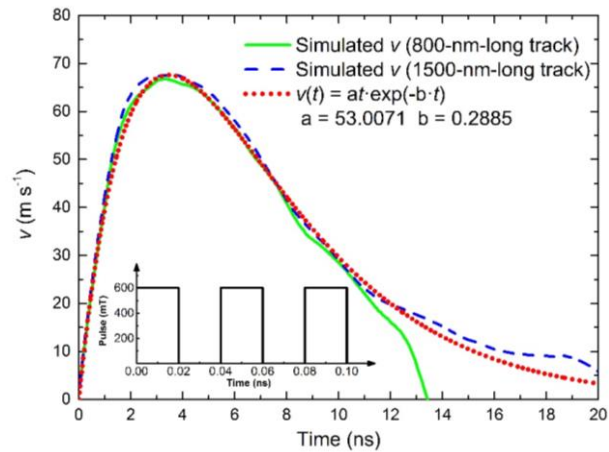

Figure 3. The velocity of a skyrmion as functions of time in 800-nm-long and 1500-nm-long nanotracks with same spin wave injection. The inset denotes the profile of the square magnetic field pulse applied along the lateral axis of the nanotrack. The red dot curve denotes the fitting function of the velocity *versus* time.

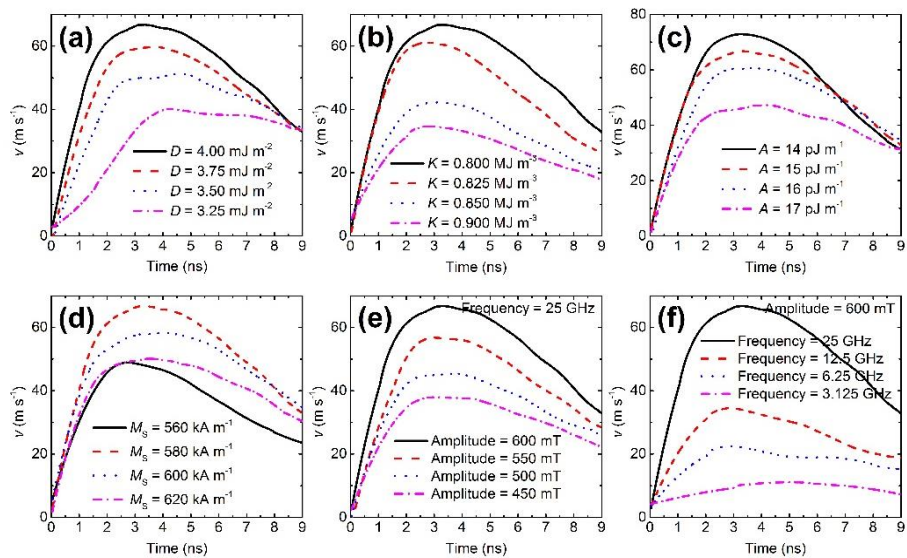

Figure 4. The velocity of a skyrmion as a function of different parameters. (a) Effect of the DMI *D*, (b) effect of the perpendicular magnetic anisotropy *K*, (c) effect of the exchange stiffness *A*, (d) effect of the saturation magnetization $M_S$, (e) effect of the amplitude of the magnetic pulse and (f) effect of the frequency of the pulse. The damping coefficient is fixed at 0.01.

It should be noted that if *D* is larger than a certain threshold (4.25 mJ m$^{-2}$ in case of Fig. 4(a)), *K* is smaller than a certain threshold (0.7 MJ m$^{-3}$ in case of Fig. 4(b)) and/or *A* is smaller than a certain threshold (13 pJ m$^{-1}$ in case of Fig. 4(c)), the skyrmion is easy to be destroyed and the system favors multiple domain walls (see Supplementary Movie 2). At the same time, if the amplitude or frequency is larger than a certain threshold, the skyrmion and/or the background magnetization of the nanotrack will be destroyed by the strong magnetic field pulse as well as the SW generated by the pulse (see Supplementary Movie 3).

In addition, we have also studied the effect of the magnetic field pulse directly on the skyrmion, *i.e.*, applied on the whole nanotrack (see Supplementary Note 4). It is found that the magnetic field pulse applied on the whole nanotrack will not drive the skyrmion into motion but may induce the breathing of the skyrmion. On the other hand, we have also investigated the motion of skyrmion driven by spin waves generated via oscillating Oersted field (see Supplementary Note 5), where the spatiotemporal dependent Oersted field acts on the whole sample but mainly focuses on the spin wave injection region, and it is found the results remain qualitatively the same with that driven by spin waves generated via pulse element.

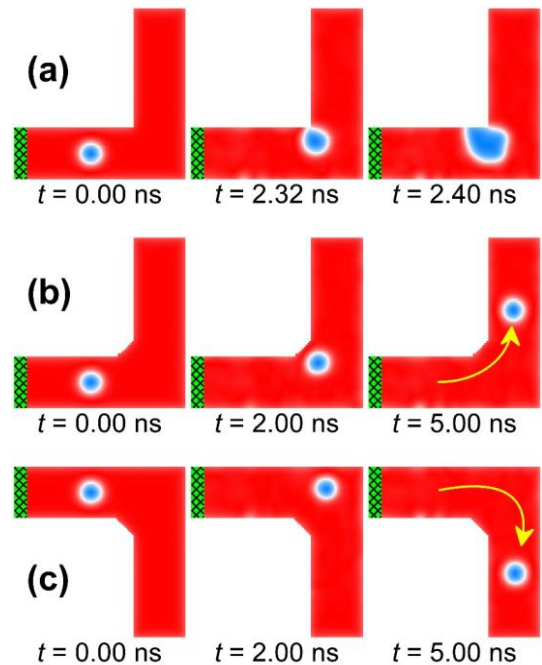

Figure 5. Snapshots of the SW-driven motion of a skyrmion ($Q = 1$) at the L-corner. The magnetic field pulse is applied along the lateral axis in the patterned green region with the profile shown in the inset of Fig. 3 (hereinafter the same). The yellow arrow denotes the motion of the skyrmion (hereinafter the same). (a) The skyrmion is destroyed by the corner due to the tilts of magnetization at the corner edge. Hence, we cut the 90-degree corner into two 135-degree corners, and the skyrmion smoothly turns left at the L-corner in (b) and turns right in (c).

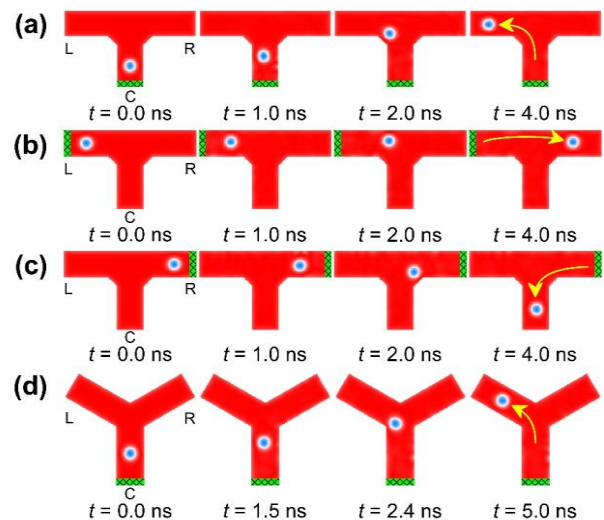

Figure 6. Snapshots of the SW-driven motion of a skyrmion ($Q = 1$) in the T-junction and Y-junction. (a) the skyrmion turns left from the C-branch into the L-branch of the T-junction. (b) the skyrmion goes straight from the L-branch to the R-branch of the T-junction. (c) the skyrmion turns left from the R-branch into the C-branch of the T-junction. (d) the skyrmion turns left from the C-branch into the L-branch of the Y-junction, similar to (a).

3.2. A skyrmion on L-corners, T- and Y-junctions

For the application of skyrmionic logic circuit, we also study the skyrmion driven by SW in constricted geometries such as L-corners, T- and Y- junctions, as shown in Fig. 1(b), 1(c) and 1(d). As shown in Fig. 5(a), the skyrmion driven by SW in a L-corner is destroyed when it turns left and touches the edge at the corner (see Supplementary Movie 4). We therefore cut the 90-degree corner into two 135-degree corners, as shown in Fig. 5(b) and 5(c). In this configuration, the skyrmion smoothly turns left without touching the edge of the L-corner in Fig. 5(b) (see Supplementary Movie 5). The spin wave can also move a skyrmion into the right direction at the L-corner as shown in Fig. 5(c) (see Supplementary Movie 6).

Figure 6 shows the SW-driven motion of the skyrmion in the T-junction and Y-junction. For the case of T-junction, we choose a skyrmion with positive topological number [29, 54]. The skyrmion always turns left, *i.e.*, from the central branch (C-branch) to the left branch (L-branch) as shown in Fig. 6(a) (see Supplementary Movie 7), from the L-branch to the right branch (R-branch) as shown in Fig. 6(b) (see Supplementary Movie 8), and from the R-branch to the C-branch as shown in Fig. 6(c) (see Supplementary Movie 9). Similarly, for the case of Y-junction, the skyrmion always turns left, as shown in Fig. 6(d) (see Supplementary Movie 10 and Supplementary Note 2). By contrast, the skyrmion always turn right if the topological number of the skyrmion is negative.

Although the SW-driven skyrmion with positive topological number on the T-junction or Y-junction has an intrinsic favor of turning left at the junction as shown in Fig. 6, it is also possible to control the turning direction of the skyrmion based on a series of SW-injection pulse elements, MTJ magnetization detectors as well as built-in circuits [32, 33, 55-57].

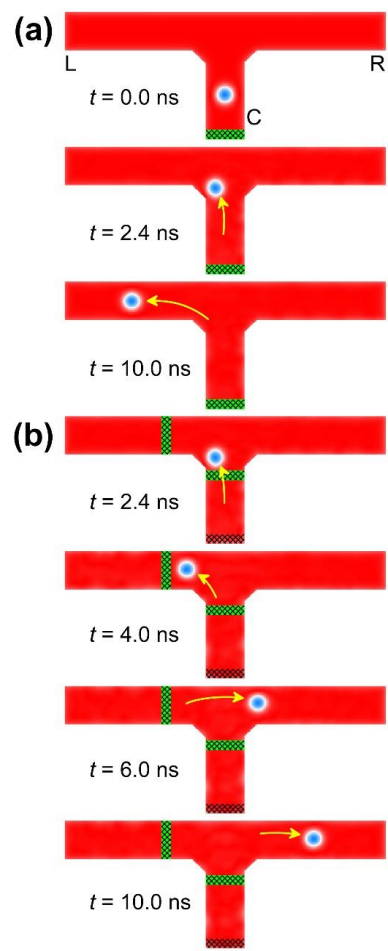

Figure 7. Control of the turning direction of the SW-driven skyrmion ($Q = 1$) at the T-junction. (a) The skyrmion naturally turns left from the C-branch into the L-branch of the T-junction. (b) The skyrmion turns right from the C-branch into the R-branch of the T-junction with the help of two magnetic pulse elements. Similar method could be applied to control the turning direction of the skyrmion in the Y-junction.

Figure 7 shows the control of the turning direction of the SW-driven skyrmion in the T-junction. As shown in Fig. 7(a), the SW-driven skyrmion with positive topological number will turn left from the C-branch into the L-branch (see Supplementary Movie 11). When the skyrmion is just out of the C-branch ($t = 2.4$ ns) as shown in Fig. 7(b), we apply two magnetic field pulses (700 mT, 25 GHz) near the exits of the C-branch and the L-branch and simultaneously switch off the pulse source (600 mT, 25 GHz) at the end of the C-branch. In this case, the skyrmion will be pushed into the R-branch by the SWs ($t = 6$ ns) (see Supplementary Movie 12).

It should be noted that all these results are valid when the direction of the magnetic pulse changes from being perpendicular to the nanotrack to being parallel to the nanotrack (see Supplementary Movies 14 – 22 and Supplementary Note 1).

On the other hand, it is worth mentioning that, due to the nonreciprocity of SW [58, 59], the SW-driven motion of the skyrmion is nonreciprocal as well. A skyrmion driven by SW from the same magnetic field pulse is different in motion depending on whether it is placed on the left or the right side of the field source in the nanotrack (see Supplementary Movie 13).

### 3.3. Thiele equation analysis

We have investigated the SW-driven skyrmion dynamics in constricted geometries including the nanotracks, L-corners, T- and Y- junctions. A skyrmion can travel in such geometries without touching the edges. It has an intrinsic tendency to turn left or right depending on the sign of the skyrmion number. By applying multiple spin wave injections, we can control the dynamics of a skyrmion. Our results will be a basis of skyrmion devices in which a sequence of skyrmions move in nano-circuits driven by spin wave.

A skyrmion is at rest initially. Once the spin wave arrives at the skyrmion, it starts the accelerated motion $\mathbf{v}^{(s)} = at$ where $a$ corresponds to the acceleration of the skyrmion. After long enough time, the velocity of skyrmion becomes the same as that of SW $\mathbf{v}^{(s)}(t) = \mathbf{v}^{(d)}(t)$. The velocity of a skyrmion also decays exponentially $\mathbf{v}^{(s)}(t) \propto e^{-bt}$ since SW decays exponentially $\mathbf{v}^{(d)}(x) = ce^{-dx}$. Accordingly we obtain the fitting function $\mathbf{v}^{(s)}(t) = ate^{-bt}$, as shown in Fig. 3. The fitting parameters are showed in Table I. We find that $a \propto c$ and $b \propto d$. The former relation implies that the initial acceleration is proportional to the amplitude of SW, while the latter relation implies the exponential decay of the velocity of a skyrmion due to the SW decay.

The acceleration $a$ is proportional to both the magnitude of the spin wave $c$ and the radius of the skyrmion (See Supplementary Note 6),

$$R_{\text{Sk}} = \frac{D\pi^2}{\frac{8}{\pi}\mu_0 H + 2K\pi}. \tag{4}$$

The skyrmion radius becomes larger with increasing $D$ and decreasing $K$. This is in good agreement with Table I, where $a$ increases with increasing $D$ and decreasing $K$. We also find that $a$ is proportional to the amplitude and the frequency of the spin wave. This is because spin wave has large energy for large amplitude and large frequency and the skyrmion radius does not change by changing the amplitude and frequency of spin wave.

For larger $K$, the spins are harder to flip, the SW propagates shorter and decays faster, resulting in a smaller speed of the skyrmion. Namely, $b$ increases with increasing $K$. On the other hand, it can be seen that $b$ is not sensitive to the other parameters in Table. I within the margin of error.

This intrinsic tendency to turn left (right) of a skyrmion with $Q = 1$ ($Q = -1$) can be understood by the Thiele equation [60-62]

$$\mathbf{G} \times \left(\mathbf{v}^{(s)} - \mathbf{v}^{(d)}\right) - \mathcal{D}\alpha\mathbf{v}^{(d)} - \mathbf{F}(\mathbf{x}) = 0, \tag{5}$$

which yields

$$-G\left(\mathbf{v}_y^{(s)} - \mathbf{v}_y^{(d)}\right) - \mathcal{D}\alpha\mathbf{v}_x^{(d)} = F_x(\mathbf{x}), \quad \mathbf{v}_x^{(s)} - \mathbf{v}_x^{(d)} - \mathcal{D}\alpha\mathbf{v}_y^{(d)} = F_y(\mathbf{x}). \tag{6}$$

They are explicitly solved as

$$\mathbf{v}_x^{(d)} = \frac{1}{G^2 + \mathcal{D}^2\alpha^2}\left[G^2\mathbf{v}_x^{(s)} - G\mathcal{D}\alpha\mathbf{v}_y^{(s)} - \mathcal{D}\alpha F_x(\mathbf{x}) - GF_y(\mathbf{x})\right], \tag{7}$$

$$\mathbf{v}_y^{(d)} = \frac{1}{G^2 + \mathcal{D}^2\alpha^2}\left[G^2\mathbf{v}_y^{(s)} + G\mathcal{D}\alpha\mathbf{v}_x^{(s)} - \mathcal{D}\alpha F_y(\mathbf{x}) + GF_x(\mathbf{x})\right], \tag{8}$$

and summarized into

$$\mathbf{v}^{(d)} = \frac{1}{1+\mathcal{D}^2\alpha^2/G^2}\mathbf{v}^{(s)} + \frac{\mathcal{D}\alpha}{G^2+\mathcal{D}^2\alpha^2}\mathbf{G}\times\mathbf{v}^{(s)} - \frac{\mathcal{D}\alpha}{G^2+\mathcal{D}^2\alpha^2}\mathbf{F}(\mathbf{x}) + \frac{1}{G^2+\mathcal{D}^2\alpha^2}(\mathbf{G}\times\mathbf{F}(\mathbf{x})). \tag{9}$$

The first term is dominant since $|G| \gg |\mathcal{D}\alpha|$. In this limit, a skyrmion moves at the same velocity as the SW i.e. $\mathbf{v}^{(d)} = \mathbf{v}^{(s)}$. However, for the higher order, the Hall effect of a skyrmion emerges. We set $\mathbf{v}_x^{(s)} \neq 0, \mathbf{v}_y^{(s)} = 0$, and $V = 0$, and get

$$\mathbf{v}_x^{(d)} = \frac{1}{1+\mathcal{D}^2\alpha^2/G^2}\mathbf{v}_x^{(s)}, \quad \mathbf{v}_y^{(d)} = \frac{-G\mathcal{D}\alpha}{G^2+\mathcal{D}^2\alpha^2}\mathbf{v}_x^{(s)}. \tag{10}$$

The Hall angle is proportional to $\alpha$. The third term $-\frac{\mathcal{D}\alpha}{G^2+\mathcal{D}^2\alpha^2}\mathbf{F}(\mathbf{x})$ in Eq. (9) represents a confining potential, while the forth term $\frac{1}{G^2+\mathcal{D}^2\alpha^2}\left(\mathbf{G}\times\mathbf{F}(\mathbf{x})\right)$ represents the motion of a skyrmion when it approaches an edge. This equation shows that a skyrmion detours the confining potential $V(\mathbf{x})$. The direction of the detour depends on the Pontryagin number since the forth term is proportional to $G$. Along the edge $\nabla V(\mathbf{x})$ is very large. Hence, the third and fourth terms are dominant over the first and the second terms. The skyrmion cannot touch the edge when $\nabla V(\mathbf{x})$ is strong enough, while it touches the edge if $\nabla V(\mathbf{x})$ is not so strong. A skyrmion detour the edge due to the fourth term.

Let us consider the case where the edge exists at $x = 0$, toward which a skyrmion is moving along the $y$ axis from the $x < 0$ side. We have $\partial_x V(\mathbf{x}) > 0$ and $\partial_y V(\mathbf{x}) = 0$. Then, a skyrmion turns left for $G = 1$ since $\mathbf{v}^{(d)} > 0$, while a skyrmion turns right for $G = -1$ since $\mathbf{v}^{(d)} < 0$.

TABLE I. Constants of the fitting functions of the velocity curves showed in Fig. 4.

| $D$ (mJ m$^{-2}$) | 3.25 | 3.50 | 3.75 | 4.00 |
|---|---|---|---|---|
| a | 21.7807 | 36.4434 | 46.7089 | 53.0071 |
| b | 0.1973 | 0.2578 | 0.2827 | 0.2885 |

| $K$ (MJ m$^{-3}$) | 0.800 | 0.825 | 0.850 | 0.900 |
|---|---|---|---|---|
| a | 53.0071 | 51.3847 | 37.3514 | 31.0713 |
| b | 0.2885 | 0.3215 | 0.3231 | 0.3267 |

| $A$ (pJ m$^{-1}$) | 14 | 15 | 16 | 17 |
|---|---|---|---|---|
| a | 58.4507 | 53.0071 | 48.7456 | 40.7391 |
| b | 0.2951 | 0.2885 | 0.2908 | 0.2948 |

| $M_S$ (kA m$^{-1}$) | 560 | 580 | 600 | 620 |
|---|---|---|---|---|
| a | 41.3331 | 53.0071 | 42.9825 | 35.3391 |
| b | 0.3161 | 0.2885 | 0.2659 | 0.2603 |

| Amplitude (mT) | 450 | 500 | 550 | 600 |
|---|---|---|---|---|
| a | 31.0865 | 37.6614 | 46.6487 | 53.0071 |
| b | 0.2939 | 0.2931 | 0.2974 | 0.2885 |

| Frequency (GHz) | 3.125 | 6.25 | 12.5 | 25 |
|---|---|---|---|---|
| a | 6.9819 | 16.1533 | 27.8073 | 53.0071 |
| b | 0.2385 | 0.2573 | 0.2889 | 0.2885 |

## 4. Conclusion

In conclusion, we have presented micromagnetic simulations and analysis that demonstrate the feasibility of spin wave-driven skyrmions in constricted geometries with the Dzyaloshinskii-Moriya interaction such as nanotracks, L-corners, T- and Y-junctions. We have found a skyrmion can turn a sharp corner without touching edges even in the case of the L-corner. A skyrmion always turns left (right) at the T- or Y-junctions when the topological number is positive (negative). Our results will pave a way to future applications to skyrmionics driven by spin wave in constricted geometries.

**Acknowledgments**

Y.Z. thanks the support by the Seed Funding Program for Basic Research and Seed Funding Program for Applied Research from the University of Hong Kong, ITF Tier 3 funding (ITS/171/13), the RGC-GRF under Grant HKU 17210014, and University Grants Committee of Hong Kong (Contract No. AoE/P-04/08). M.E. thanks the support by the Grants-in-Aid for Scientific Research from the Ministry of Education, Science, Sports and Culture, No. 25400317.

Y.W.L. thanks the support by the National Natural Science Foundation of China (Grant Nos. 10974142, 51471118). G.P.Z. thanks the support by the National Natural Science Foundation of China (Grant Nos. 11074179, 10747007). M.E. is very much grateful to N. Nagaosa and J. Iwasaki for many helpful discussions on the subject. X.C.Z. thanks M. Beg, J. Iwasaki and J. Xia for useful discussions on this work.